\documentstyle[11pt]{article}

\begin{document}

\renewcommand{\thesection}{\Roman{section}}

\renewcommand{\theequation}{\arabic{equation}}
\renewcommand{\thepage}{\arabic{page}}

\def\BM#1{\mbox{\boldmath{$#1$}}}
\def\fnote#1#2{\begingroup\def\thefootnote{#1}\footnote{#2}\addtocounter
{footnote}{-1}\endgroup}

\hfill{UTTG-16-12}

\vspace{36pt}

\begin{center}
{\large {\bf {Minimal Fields of Canonical Dimensionality are Free}}}

\vspace{36pt}
Steven Weinberg\fnote{*}{Electronic address:
weinberg@physics.utexas.edu}\\
{\em Theory Group, Department of Physics, University of
Texas\\
Austin, TX, 78712}

\vspace{30pt}

\noindent
{\bf Abstract}
\end{center}

\noindent
It is shown that in a scale-invariant relativistic field theory, any field $\psi_n$ belonging to the $(j,0)$ or $(0,j)$ representations of the Lorentz group and with dimensionality $d=j+1$ is a free field.  For other field types there is no value of the dimensionality that guarantees that the field is free.  Conformal invariance is not used in the proof of these results, but it gives them a special interest; as already known and as shown here in an appendix, the only fields in a conformal field theory that can describe massless particles belong to the $(j,0)$ or $(0,j)$ representations of the Lorentz group and have dimensionality $d=j+1$.  Hence in conformal field theories massless particles are free.

\vfill

\pagebreak


  This note will show   that in a scale-invariant relativistic field theory, any  fields that belong to the minimal $2j+1$-component $(j,0)$ or $(0,j)$   representations of the Lorentz group (where $j$ is an integer or half-integer) and have canonical dimensionality  $d=j+1$  are necessarily  free fields.   This conclusion is already known for $j=0$ [1]; here it is extended to all spins.  Although conformal invariance is not used here, this  result gains interest from the fact [2] that in conformal field theories the only fields that can describe massless particles belong to the  $(j,0)$ and $(0,j)$ representations of the Lorentz group and have canonical dimensionality.  An  elementary proof of this theorem is given in an appendix.  It follows that, according to the main result of the present paper, massless particles in a conformally invariant field theory must be free particles.

To begin, consider a field $\psi_n(x)$ belonging to any representation of the Lorentz group.  Poincar\'{e} invariance tells us that 
\begin{equation}
{\cal L}_{\rho\sigma}G_{nm}(z)=-\sum_l [{\cal J}_{\rho\sigma}]_{nl}G_{lm}(z)+\sum_l G_{nl}(z)[{\cal J}^\dagger_{\rho\sigma}]_{lm}
\end{equation}
where  $G$  is the vacuum expectation value
\begin{equation}
G_{nm}(x-y)\equiv \Big\langle 0 \Big|\, \psi_n(x)\,\psi_m^\dagger(y)\,\Big|0\Big\rangle\;,
\end{equation}
${\cal L}_{\rho\sigma}$ are the differential operators
\begin{equation}
{\cal L}_{\rho\sigma}\equiv -i z^\rho\frac{\partial}{\partial z_\sigma}+i
z^\sigma\frac{\partial}{\partial z_\rho}\;,
\end{equation}
and $[{\cal J}_{\rho\sigma}]_{nm}$ are the matrices representing the generators of the Lorentz group in the representation furnished by the field $\psi(x)$.  
Iteration of Eq.~(1) gives (suppressing matrix indices)
\begin{equation}
{\cal L}^{\rho\sigma}{\cal L}_{\rho\sigma}G(z)={\cal J}^{\rho\sigma}{\cal J}_{\rho\sigma}G+
G{\cal J}^{\dagger\;\rho\sigma}{\cal J}^\dagger_{\rho\sigma}-2{\cal J}^{\rho\sigma}G{\cal J}^\dagger_{\rho\sigma}
\end{equation}

The point of this exercise is that by  elementary commutations of derivatives and coordinates, one can derive the identity
\begin{equation}
{\cal L}^{\rho\sigma}{\cal L}_{\rho\sigma}=-2z^2\Box+2S^2-4S\;,
\end{equation}
where $\Box\equiv \partial^2/\partial z^\rho\partial z_\rho$ is the usual d'Alembertian, and $S$ is the scale transformation operator
\begin{equation}
S\equiv -z^\rho \frac{\partial}{\partial z^\rho}\;.
\end{equation}
(This is analogous to the identity in three dimensions that can be used to show that the Laplacian of the spherical polynomial $r^\ell Y_\ell^m(\theta,\phi)$ vanishes.)
We will use Eqs.~(4)--(6) to show that if $\psi$ belongs to the
$(j,0)$ or $(0,j)$   representations of the Lorentz group and has canonical dimensionality then $\Box \psi=0$.

If $\psi(x)$ belongs to the $(j,0)$ representation of the Lorentz group, then 
\begin{equation}
{\cal J}_{ij}=\epsilon_{ijk}{\cal J}_k\;,~~~~{\cal J}_{i0}=-i{\cal J}_i\;,
\end{equation}
where ${\cal J}_i$ are the Hermitian matrices representing the generators of the rotation group in its spin $j$ representation.  It follows that 
\begin{eqnarray}
&&\frac{1}{2}{\cal J}^{\rho\sigma}{\cal J}_{\rho\sigma}G=2{\cal J}_i{\cal J}_iG=2j(j+1)G\nonumber\\
&&\frac{1}{2}G{\cal J}^{\dagger\rho\sigma}{\cal J}^\dagger_{\rho\sigma}=2G{\cal J}^\dagger_i{\cal J}^\dagger_i=2j(j+1)G\\
&&{\cal J}^{\rho\sigma}G{\cal J}^\dagger_{\rho\sigma}=0\nonumber\;.
\end{eqnarray}
Also, if $\psi$ has dimensionality $d$ (counting powers of momentum)  then in a scale-invariant theory
\begin{equation}
S\,G(z)=2dG(z)\;.
\end{equation}
So for these fields, Eq.~(4) reads
\begin{equation}
-2z^2\Box G(z)+8d^2\,G(z)-8d\,G(z)=8j(j+1)\,G(z)\;,
\end{equation}
and in particular, for $d=j+1$,
\begin{equation}
\Box G(z)=0\;.
\end{equation}

Operating again with a d'Alembertian, it follows trivially that
\begin{equation}
0=\Big\langle 0 \Big|\, \Box_x\psi_n(x)\,[\Box_y\psi_m(y)]^\dagger\,\Big|0\Big\rangle\;,
\end{equation}
so 
\begin{equation}
[\Box_y\psi_m(y)]^\dagger\,\Big|0\Big\rangle=0
\end{equation}
But any local operator that annihilates the vacuum must vanish[3], so 
\begin{equation}
\Box_y\psi_m(y)=0\;,
\end{equation}
and the field is therefore free.  The proof for $(0,j)$ fields is identical, except for an inconsequential difference of sign of ${\cal J}_{i0}$.  This does not say that the theory for which 
$\Box_y\psi_m(y)=0$ is a free-field theory, but only that the field $\psi_n$ is free; there may be other fields in the same theory, which transform according to other representations of the Lorentz group and/or have other dimensionalities, that are not free.

It is only  fields belonging to the $(j,0)$ or $(0,j)$ representations of the Lorentz group that can be shown in this way to be free.  Indeed, for fields $\chi_r$ belonging to other irreducible representations of the Lorentz group, there is {\em no} value of  dimensionality $d$ for which it is guaranteed that $\Box \chi_r=0$.  If $\chi_r$ transforms according to the $(j,j')$ representation, then $\chi_r\chi_s^\dagger$ in general transforms reducibly, as a sum of the representations $(A,B)$ with both $A$ and $B$ running by unit steps from $|j-j'|$ to $j+j'$.  The vacuum expectation value $F\equiv\langle 0|\chi_r\chi_s^\dagger|0\rangle$  has a corresponding decomposition into terms $F^{(A,B)}$ belonging to the same representations, for which in place of Eq.~(10) we have
\begin{equation}
-2z^2\Box F^{(A,B)}(z)+8d^2\,F^{(A,B)}(z)-8d\,F^{(A,B)}(z)=4[A(A+1)+B(B+1)]\,F^{(A,B)}(z)\;.
\end{equation}
If $F^{(A,B)}$ itself is non-zero, then the only way that $\Box F^{(A,B)}$ can vanish is if $d$ takes a value for which 
\begin{equation}
2d(d-1)=A(A+1)+B(B+1)\;.
\end{equation}
There is obviously no value of $d$ for which this is satisfied  for all values of $A$ and $B$ between $|j-j'|$ and $j+j'$ unless either $j'=0$ or $j=0$, in which case both $A$ and $B$ take the unique value $j$ or $j'$.

\vspace{15pt}

\begin{center}
{\bf ACKNOWLEDGMENTS}
\end{center}

I am grateful for helpful conversations and correspondence with J. Distler, H. Osborn, E. Sezgin, and E. Witten.
  This material is based upon work supported by the National Science Foundation under Grant Number PHY-0969020 and with support from The Robert A. Welch Foundation, Grant No. F-0014.

\vspace{15pt}

\begin{center}
{\bf APPENDIX: MASSLESS PARTICLE FIELDS IN CONFORMAL THEORIES}

\end{center}

\setcounter{equation}{0}

\renewcommand{\theequation}{A.\arabic{equation}}

This appendix offers an elementary demonstration of Mack's result [2], that the only fields in a conformal field theory that can describe a massless particle of helicity $j$ or $-j$ (in the sense that the field has a non-vanishing matrix element between the particle state and the vacuum) is a $(0,j)$ or $(j,0)$ field of canonical dimensionality $d=j+1$.  Together with the main result of the present work, this shows that massless particles in conformal field theories are free.

It is necessary first to say how massless particle states transform under infinitesimal conformal transformations.  This is already known[4] (as I learned after working out the transformation rules), but it is worth presenting a detailed derivation here to show that these transformation rules are unique.      To define the massless particle  states, we first introduce a standard three-momentum $\kappa\hat{z}$ of magnitude $ \kappa$ in the $+3$-direction, and define a state
$|\kappa\hat{z},\sigma\rangle$ with this momentum and with helicity $\sigma$, in the sense that this is an eigenstate of the generator $J_{12}$ of rotations in the $1-2$ plane with eigenvalue $\sigma$:
\begin{equation}
 J_{12}|\kappa\hat{z},\sigma\rangle=\sigma|\kappa\hat{z},\sigma\rangle\;.
\end{equation}
In order to avoid introducing new continuous degrees of freedom, it is also necessary to assume that these states are annihilated by the generators of the invariant Abelian subgroup of the little group (the group of Lorentz transformations that leave the standard three-momentum invariant):
\begin{equation}
[J_{10}+J_{13}]|\kappa\hat{z},\sigma\rangle=[J_{20}+J_{23}]|\kappa\hat{z},\sigma\rangle=0\;.
\end{equation}
We then take
\begin{equation}
|{\bf p},\sigma\rangle \equiv U\Big(L({\bf p})\Big)|\kappa\hat{z},\sigma\rangle\;,
\end{equation}
where $U\Big(L({\bf p})\Big)$  is the unitary operator representing a standard Lorentz transformation $L^{\mu}{}_\nu({\bf p})$ that takes the standard three-momentum $\kappa\hat{z}$ to ${\bf p}$.  
For instance, we can take $L^{\mu}{}_\nu({\bf p})$ as a boost along the 3-axis that takes 
$\kappa\hat{z}$ to $|{\bf p}|\hat{z}$, followed by a rotation in the $\hat{z}-\hat{p}$ plane that takes 
$|{\bf p}|\hat{z}$ to ${\bf p}$.  These states are here normalized to have the Lorentz-invariant scalar product
\begin{equation}
\langle {\bf p},\sigma|{\bf p}',\sigma'\rangle =\delta_{\sigma\sigma'}|{\bf p}|\,\delta^3({\bf p}-{\bf p}')\;,
\end{equation}
rather than the conventional scalar product which does not contain the factor $|{\bf p}|$.
 As is well known[5], acting on such a state, the unitary operator $U(\Lambda)$ representing a Lorentz transformation $\Lambda^\mu{}_\nu$ gives\fnote{**}{We take $i,\,j,\,k,\cdots$ to run over the spatial coordinate indices $1,\,2,\,3$, while $\mu,\nu,\,\cdots$ run over the spacetime indices $0,\,1,\,2,\,3$.  Repeated indices are summed, and the spacetime metric $\eta_{\mu\nu}$ has non-zero components $\eta_{11}=\eta_{22}=\eta_{33}=1$, $\eta_{00}=-1$.}
\begin{equation}
U(\Lambda)|{\bf p},\sigma\rangle=\exp\Big(i\sigma \phi({\bf p},\Lambda)\Big)|{\bf \Lambda p},\sigma\rangle
\end{equation}
where $(\Lambda p)^i\equiv \Lambda^i{}_j p^j+\Lambda^i{}_0|{\bf p}|$,
and $\phi$ is the real angle appearing in the Wigner rotation for massless particles: 
\begin{equation}
U\Big(L^{-1}({\bf \Lambda p})\Lambda\L({\bf p})\Big)|\kappa\hat{z},\sigma\rangle=
\exp\Big(i\sigma \phi({\bf p},\Lambda)\Big)|\kappa\hat{z},\sigma\rangle\;.
\end{equation}

For infinitesimal Lorentz transformations $\Lambda^\mu{}_\nu=\delta^\mu{}_\nu+\omega^\mu{}_\nu$ (with $\omega^{\mu\nu}\equiv \omega^\mu{}_\lambda\,\eta^{\nu\lambda}$ infinitesimal and antisymmetric) we have
\begin{equation}
U(1+\omega)=1+\frac{i}{2}\omega^{\mu\nu}J_{\mu\nu}
\end{equation}
with $J_{\mu\nu}$ Hermitian and antisymmetric in $\mu$ and $\nu$, and satisfying the commutation relations
\begin{equation}
i[J_{\mu\nu},J_{\kappa\lambda}]=\eta_{\nu\kappa}J_{\mu\lambda}-\eta_{\mu\kappa}J_{\nu\lambda}-\eta_{\nu\lambda}J_{\mu\kappa}+\eta_{\mu\lambda}J_{\nu\kappa}\;.
\end{equation}
In this case, Eq.~(A.5) reads
\begin{eqnarray}
J_{ij}|{\bf p},\sigma\rangle &=&\left[i\left(p_i\frac{\partial}{\partial p_j}-p_j\frac{\partial}{\partial p_i}\right)+\sigma\phi_{ij}({\bf p})\right]|{\bf p},\sigma\rangle\;,\\
J_{i0}|{\bf p},\sigma\rangle &=&\left[i|{\bf p}|\frac{\partial}{\partial p_i}+\sigma\phi_{i0}({\bf p})\right]|{\bf p},\sigma\rangle\;,
\end{eqnarray}
where the antisymmetric real coefficients $\phi_{\mu\nu}$ (which will play a large role in what follows) are defined by
\begin{equation}
\phi({\bf p},1+\omega)=\frac{1}{2}\omega^{\mu\nu}\phi_{\mu\nu}({\bf p})\;.
\end{equation}
Of course, also
\begin{equation}
P^\mu|{\bf p},\sigma\rangle=p^\mu|{\bf p},\sigma\rangle\;,
\end{equation}
with $ p^0=|{\bf p}|$.
It is straightforward to check that the operators (A.9), (A.10), and (A.12) are Hermitian within the norm (A.4).

The functions $\phi_{\mu\nu}$ can be calculated from Eq.~(A.6) in any convenient representation of the Lorentz group, such as the two-component fundamental spinor representation.  The result is
\begin{equation}
\phi_{ij}({\bf p})=\frac{\epsilon_{ijk}\,(\hat{p}+\hat{z})_k}{1+\hat{p}\cdot\hat{z}}\;,~~~~
\phi_{i0}({\bf p})=-\frac{(\hat{p}\times \hat{z})_i}{1+\hat{p}\cdot\hat{z}}\;.
\end{equation}
These formulas  depend on a particular prescription for  the standard Lorentz transformation $L({\bf p})$ that takes $\kappa \hat{z}$ to ${\bf p}$.  Suppose we change this prescription by introducing some other Lorentz transformation  $L'({\bf p})$ that takes $\kappa \hat{z}$ to ${\bf p}$,  The Lorentz transformation $L^{-1}({\bf p})L'({\bf p})$ is an element of the little group, and therefore merely muliples $|\kappa\hat{z},\sigma\rangle$ with a phase factor $\exp\Big(i\zeta({\bf p})\Big)$, so if we use $L'({\bf p})$ in place of $L({\bf p})$ in Eq.~(3) the one particle state is changed to 
\begin{eqnarray}
&&|{\bf p},\sigma\rangle' \equiv U\Big(L'({\bf p})\Big)|\kappa\hat{z},\sigma\rangle=U\Big(L({\bf p})\Big)U\Big(L^{-1}({\bf p})L'({\bf p})\Big)|\kappa\hat{z},\sigma\rangle\nonumber\\&& =\exp\Big(i\zeta({\bf p})\Big)|{\bf p},\sigma\rangle
\;.
\end{eqnarray}
 Thus the formulas (A.13) for $\phi_{ij}({\bf p})$ and $\phi_{i0}({\bf p})$ represent a particular convention for the phase of these states, supplementing  the convention (A.4) we have adopted for the normalization of the states.

The generators of the conformal symmetry group in four spacetime dimensions comprise the generator $S$ of scale transformations and the generators $K_\mu$ of special conformal transformations, together with the generators $J_{\mu\nu}$ and $P_\mu$ of Poincar\'{e} transformations.  They satisfy the well-known commutation relations
\begin{eqnarray}
&&i\left [ K_\mu\;,\; J_{\rho\lambda}\right] = \eta_{\mu\rho}
K_\lambda -
\eta_{\mu\lambda} K_\rho\;,\\
&&[K_\mu,K_\nu]=0\;,\\
&& i[P_\mu\,,\,K_\nu]=2J_{\mu\nu}+2\eta_{\mu\nu}S\;,\\
&& i[S,P_\mu]=P_\mu\;,\\
&& i[S,K_\mu]=-K_\mu\;,\\
&&[S,J_{\mu\nu}]=0\;,
\end{eqnarray}
as well as the familiar commutation relations of the Poincar\'{e} group:
\begin{eqnarray}
& & i\left[ J_{\mu\nu},J_{\rho\lambda}\right] = \eta_{\nu\rho}
J_{\mu\lambda} -
\eta_{\mu\rho} J_{\nu\lambda} -\eta_{\lambda\mu} J_{\rho\nu} +
\eta_{\lambda\nu}
J_{\rho\mu} \;, \\
& & i\left [ P_\mu\;,\; J_{\rho\lambda}\right] = \eta_{\mu\rho}
P_\lambda -
\eta_{\mu\lambda} P_\rho\;, \\
& & ~\left[P_\mu\;,\;P_\rho\right] = 0\;.
\end{eqnarray}
It is straightforward though quite tedious to show that these commutation relations are satisfied by the following operators on one-particle states:
\begin{eqnarray}
&&K_0|{\bf p},\sigma\rangle=\Bigg[|{\bf p}|\frac{\partial^2}{\partial p_k\partial p_k}-2i\sigma\phi_{k0}({\bf p})\frac{\partial}{\partial p_k}\nonumber\\&&~~~~~~~~~~~~~-\frac{\sigma^2}{|{\bf p}|}\left(\phi_{k0}({\bf p})\phi_{k0}({\bf p})+1\right)\Bigg]|{\bf p},\sigma\rangle\\
&&K_i|{\bf p},\sigma\rangle=\Bigg[2p_k\frac{\partial^2}{\partial p_k\,\partial p_i}-p_i \frac{\partial^2}{\partial p_k\,\partial p_k}+2\frac{\partial}{\partial p_i}\nonumber\\&&~~~~+2i\sigma\phi_{ik}({\bf p})\frac{\partial}{\partial p_k}+\frac{\sigma^2}{|{\bf p}|}\Big(2\phi_{ik}({\bf p})\phi_{k0}({\bf p})-\hat{p}_i[\phi_{k0}({\bf p})\phi_{k0}({\bf p})+1]\Big)\Bigg]|{\bf p},\sigma\rangle\;,\nonumber \\&&{}\\
&&S|{\bf p},\sigma\rangle=i\left[p_k\frac{\partial}{\partial p_k}+1\right]|{\bf p},\sigma\rangle\;,
\end{eqnarray}
together with the Poincar\'{e} transformation operators (A.9), (A.10), and (A.12).  (These results agree with those of [4] when we use the formulas (A.13) for $\phi_{\mu\nu}({\bf p})$, but in what follows it will be convenient to leave the transformation rules in the form (A.24)--(A.26).)
  What is less straightforward is to show that, given massless particle states $|{\bf p},\sigma\rangle$  satisfying the Poincar\'{e} transformation rules (A.9), (A.10), and (A.12),  the only conformal transformation properties consistent with the commutation relations are those given in Eqs.~(A.24)--(A.26).  To show this, we will outline the steps by which   
Eqs.~(A.24)--(A.26) are derived.

\vspace{10pt}
\noindent
(a) The commutation relation (A.17) gives $[K_0,P_i]=2iJ_{i0}$.  This uniquely fixes the derivative terms in Eq.~(A.24).  That is, 
\begin{equation}
 K_0|{\bf p},\sigma\rangle=\Bigg[|{\bf p}|\frac{\partial^2}{\partial p_k\partial p_k}-2i\sigma\phi_{k0}({\bf p})\frac{\partial}{\partial p_k}+\alpha({\bf p})\Bigg]|{\bf p},\sigma\rangle\;,
\end{equation}
with $\alpha({\bf p})$ some c-number function that remains to be calculated.  Using  Eqs.~(A.4) and (A.13), we easily see that for $K_0$ to be Hermitian,  $\alpha({\bf p})$ must be real.  

\vspace{10pt}
\noindent
(b) The commutation relation (A.17) also gives $[K_0,P_0]=-2iS$.  Using Eqs.~(A.12) and (A.27), we obtain Eq.~(A.26) for the action of the dilation generator $S$ on one-particle states.

 \vspace{10pt}
\noindent
(c) Using Eq.~(A.13) together with Eqs.~(A.27) and (A.9),  we can calculate that
$$ [J_{ij},K_0]|{\bf p},\sigma\rangle=-i|{\bf p},\sigma\rangle\left\{p_i\frac{\partial}{\partial p_j}-p_j\frac{\partial}{\partial p_i}\right\}\Big[\alpha({\bf p})+\phi_{k0}({\bf p})\phi_{k0}({\bf p})/|{\bf p}|\Big]\;.
$$
Since this must vanish, $\alpha({\bf p})+\phi_{k0}({\bf p})\phi_{k0}({\bf p})/|{\bf p}|$ must be a function only of the modulus $|{\bf p}|$.  
Further,  Eqs.~(A.19) and (A.26) tell us that this function scales as $1/p$, and hence must take the form $a/|{\bf p}|$, with $a$  real and ${\bf p}$-independent.  Hence   Eq.~(A.27) reads
\begin{equation}
 K_0|{\bf p},\sigma\rangle=\Bigg[|{\bf p}|\frac{\partial^2}{\partial p_k\partial p_k}-2i\sigma\phi_{k0}({\bf p})\frac{\partial}{\partial p_k}-\frac{\phi_{k0}({\bf p})\phi_{k0}({\bf p})-a}{|{\bf p}|} \Bigg]|{\bf p},\sigma\rangle\;,
\end{equation}

 \vspace{10pt}
\noindent
(d) From Eq.~(A.15), we have $K_i=-i[J_{i0},K_0]$.  Using the formulas (A.10) and (A.28) for the operators $J_{i0}$ and $K_0$ , we find
\begin{eqnarray}
 && K_i|{\bf p},\sigma\rangle=\Bigg[2p_k\frac{\partial^2}{\partial p_i\partial p_k}-p_i\frac{\partial^2}{\partial p_k\,\partial p_k}+2\frac{\partial}{\partial p_i}+2i\sigma\phi_{ik}({\bf p})\frac{\partial}{\partial p_k}\nonumber\\&&~~~
-\frac{2\sigma^2}{|{\bf p}|}\phi_{k0}({\bf p})\phi_{ki}({\bf p})-\hat{p}_i\frac{\sigma^2\phi_{k0}({\bf p})\phi_{k0}({\bf p})-a}{|{\bf p}|}\Bigg]|{\bf p},\sigma\rangle\;.
\end{eqnarray}

 \vspace{10pt}
\noindent
(e) It only remains to find the constant $a$.  We can do this by requiring that $K_\mu$ be a four-vector.  Since we have constructed $K_0$ to be a rotational scalar and $K_i$ to be equal to $-i[J_{i0},K_0]$, the remaining requirement provided by Eq.~(A.15) is that $[J_{i0},K_j]=i\delta_{ij}K_0$. Now using the formulas (A.13), equating the coefficients of $\delta_{ij}$ in the non-derivative terms on both sides of this commutation relation, we find $a=-\sigma^2$.  Eqs.~(A.28) and (A.29) are then the desired results (A.24) and (A.25).  

\vspace{10pt}

Now let us turn to the transformation of field operators.  We will consider here only fields that transform linearly and homogeneously under Poincar\'{e} transformations:
\begin{equation}
i[J^{\mu\nu},\psi_n(x)]=-i\sum_m{\cal J}^{\mu\nu}_{nm}\psi_m(x)+\left(x^\nu\partial^\mu-x^\mu\partial^\nu\right)\psi_n(x)
\end{equation}
\begin{equation}
i[P_\mu,\psi_n(x)]=-\partial_\mu\psi_n(x)
\end{equation}
where ${\cal J}^{\mu\nu}$  is a set of spin matrices that satisfy the same commutation relation (A.21) as $J^{\mu\nu}$.
This excludes gauge fields, whose Lorentz transformation properties in an operator formalism (rather than a path-integral formalism) include a gauge transformation in addition to the transformation (A.30).

A primary field may be defined as one with the familiar conformal transformation properties:
\begin{equation}
i[K_\nu,\psi_n(x)]=-2i\sum_m[{\cal J}_{\mu\nu}]_{nm}x^\mu\psi_m(x)+2d_n\,x_\nu\psi_n(x)+(2x_\nu x^\rho\partial_\rho-x^2\partial_\nu)\psi_n(x)\;,
\end{equation}
\begin{equation}
i[S,\psi_n(x)]=d_n\psi_n(x)+x^\mu\partial_\mu\psi_n(x)\;,
\end{equation}
where $d_n$ is a real number, known as the conformal dimensionality of the field.  Since neither Lorentz nor conformal transformations mix different irreducible representations of the Lorentz group, we will assume without loss of generality that the matrices ${\cal J}_{\mu\nu}$ furnish an irreducible representation of the algebra of the Lorentz group.

Our aim in this appendix is to find what kinds of primary fields can describe a massless particle of a given helicity $\sigma$.  By a field ``describing'' a particle, we mean that the field has non-vanishing matrix elements between the particle state and the vacuum.  The propagator of such a field will have a zero mass pole whose residue is proportional to the product of this matrix element and its complex conjugate, so that S-matrix elements for this particle can be found from the residues of poles in the vacuum expectation value of time-ordered products of the field.

Let's first take up the simple case of dilations.  Assuming that the vacuum is invariant under these transformations, Eq.~(A.33) gives
\begin{equation}
-i\langle 0 | \psi_n(0) S |{\bf p},\sigma\rangle=d_n\langle 0 | \psi_n(0)  |{\bf p},\sigma\rangle 
\end{equation}
With Eq.~(A.26), this becomes
\begin{equation}
d_n\langle 0 | \psi_n(0)  |{\bf p},\sigma\rangle=\left[p_k\frac{\partial}{\partial p_k}+1\right]\langle 0 | \psi_n(0)  |{\bf p},\sigma\rangle \;.
\end{equation}
The way that the matrix element $\langle 0 | \psi_n(0)  |{\bf p},\sigma\rangle$ scales with momentum depends on the Lorentz transformation properties of  the field $\psi_n$.  Recall that 
the general irreducible representations of the Lorentz group are labeled $(A,B)$, where $A$ and $B$ are positive integers or half-integers.  These representations are defined by writing the  matrices representing the generators of the Lorentz group in terms of two Hermitian  matrix 3-vectors defined by
\begin{equation}
{\cal A}_i\equiv \frac{1}{2}{\cal J}_i+\frac{i}{2}{\cal J}_{i0}\;,~~~
{\cal B}_i\equiv \frac{1}{2}{\cal J}_i-\frac{i}{2}{\cal J}_{i0}\;,
\end{equation}
where as usual ${\cal J}_i\equiv \frac{1}{2}\epsilon_{ijk}{\cal J}_{jk}$.
The commutation relations of the ${\cal J}_{\mu\nu}$ tell us that
\begin{equation}
[{\cal A}_i,{\cal A}_j]=i\epsilon_{ijk}{\cal A}_k\;,~~~~~[{\cal B}_i,{\cal B}_j]=i\epsilon_{ijk}{\cal B}_k\;,~~~~[{\cal A}_i,{\cal B}_j]=0\;.
\end{equation}
In the $(A,B)$ representation of the Lorentz group,  these are $A\times A$ and $B\times B$ matrices,   such that
\begin{equation}
\BM{\cal A}^2=A(A+1)\,,~~~~~~~\BM{\cal B}^2=B(B+1)\;.
\end{equation}
It is an old result [6] that the only free fields that can describe a massless particle of helicity $\sigma$ have 
\begin{equation}
\sigma = B-A
\end{equation}
and have matrix  elements between the vacuum and these one-particle states that scale as $p^{A+B}$; that is
\begin{equation}
p_k\frac{\partial}{\partial p_k}\langle 0 | \psi_n(0)  |{\bf p},\sigma\rangle=
(A+B)\langle 0 | \psi_n(0)  |{\bf p},\sigma\rangle\;.
\end{equation}
It is easy to show using Lorentz invariance that Eqs.~(A.39) and (A.40)  also hold for general interacting fields.  Combining Eq.~(A.40)  with Eq.~(A.35) gives the conformal dimensionality
\begin{equation}
d=A+B+1\;.
\end{equation}
in which we drop the subscript $n$ on $d_n$ since this is the same for all components of a field belonging to an irreducible representation of the Lorentz algebra.
Thus   fields of Lorentz type $(A,B)$ that describe a massless particle (in the sense explained above) can only have conformal dimensionality $A+B+1$

Now let's consider special conformal transformations.  We will compare what we have learned  about the conformal transformation properties of one-particle states with the consequences of the conformal transformation properties of a primary field that can describe such a particle.  By taking the matrix element of Eq.~(A.32) for $\nu=0$ between a one-particle state $|{\bf p},\sigma\rangle$ and the vacuum, and assuming that the vacuum is conformal-invariant, we find
\begin{eqnarray}
&& \langle 0 |\psi_n(x)K_0|{\bf p},\sigma\rangle=2\sum_n[{\cal J}_{i0}]_{nm}x^i\langle 0 |\psi_m(x)|{\bf p},\sigma\rangle \nonumber\\&&~~~
+2id\,x_0\langle 0 |\psi_n(x)|{\bf p},\sigma\rangle-(2x_0 x^\rho p_\rho+x^2)|{\bf p}|\langle 0 |\psi_n(x)|{\bf p},\sigma\rangle\;.~~~~~~
\end{eqnarray}
We need to rearrange the right-hand side of  Eq.~(A.42) so that it takes the form $\langle 0 |\psi_n(x){\cal K}|{\bf p},\sigma\rangle$, where ${\cal K}$ is some $x^\rho$-independent matrix function of momentum and momentum derivatives, and then compare ${\cal K}|{\bf p},\sigma\rangle$ with what Eq.~(A.24) gives for $K_0|{\bf p},\sigma\rangle$.  For this purpose, we first re-write Eq.~(A.42) so that it reads
\begin{eqnarray}
&& \langle 0 |\psi_n(x)K_0|{\bf p},\sigma\rangle=2\sum_n[{\cal J}_{i0}]_{nm}x^i\langle 0 |\psi_m(x)|{\bf p},\sigma\rangle \nonumber\\&&~~~
+2i(d\,-1)x_0\langle 0 |\psi_n(x)|{\bf p},\sigma\rangle+\langle 0 |\psi_n(0)|{\bf p},\sigma\rangle\,|{\bf p}|\frac{\partial^2 e^{ip\cdot x}}{\partial p_k\partial p_k}\nonumber\\&&=
2\sum_n[{\cal J}_{i0}]_{nm}x^i\langle 0 |\psi_m(x)|{\bf p},\sigma\rangle+2ix_0\,e^{ip\cdot x}\left[d\,-1-p_k\frac{\partial}{\partial p_k}\right]\langle 0 |\psi_n(0)|{\bf p},\sigma\rangle\nonumber\\&&
~~~-2i|{\bf p}|e^{ip\cdot x}x_k\frac{\partial}{\partial p_k}\langle 0 |\psi_n(0)|{\bf p},\sigma\rangle-|{\bf p}|e^{ip\cdot x}\frac{\partial^2}{\partial p_k\partial p_k}\langle 0 |\psi_n(0)|{\bf p},\sigma\rangle\nonumber\\&&~~~+\langle 0 |\psi_n(x)|{\bf p}|\frac{\partial^2}{\partial p_k\partial p_k}|{\bf p},\sigma\rangle\;.
\end{eqnarray} 
Eq.~(A.35) tells us that the term in the final expression proportional to $x_0$ vanishes, as it must.  

To calculate the derivatives of $\langle 0 |\psi_n(0)|{\bf p},\sigma\rangle$ with respect to momentum, we use Lorentz invariance.  Combining Eq.~(A.30) for $x=0$ with Eq.~(A.10), we find
$$ i|{\bf p}|\frac{\partial}{\partial p_k}\langle 0 |\psi_n(0)|{\bf p},\sigma\rangle=\sum_m[{\cal J}_{k0}]_{nm}\langle 0 |\psi_m(0)|{\bf p},\sigma\rangle-\sigma\phi_{k0}\langle 0 |\psi_n(0)|{\bf p},\sigma\rangle\;.$$
Using this formula, and its derivative with respect to $p_k$, together with Eq.~(A.13), we put Eq.~(A.43) in the form
\begin{eqnarray}
&& \langle 0 |\psi_n(x)K_0|{\bf p},\sigma\rangle=\frac{1}{|{\bf p}|}\sum_m[{\cal J}_{k0}{\cal J}_{k0}]_{nm}
\langle 0 |\psi_m(x)|{\bf p},\sigma\rangle \nonumber\\&&~~~
-\frac{2\sigma\phi_{k0}}{|{\bf p}|}\sum_m[{\cal J}_{k0}]_{nm}
\langle 0 |\psi_m(x)|{\bf p},\sigma\rangle\nonumber\\&&
+\sigma^2\phi_{k0}\phi_{k0}\langle 0 |\psi_n(x)|{\bf p},\sigma\rangle
+2\sigma\phi_{k0}x_k\langle 0 |\psi_n(x)|{\bf p},\sigma\rangle\nonumber\\&&~~~
+\langle 0 |\psi_n(x)|{\bf p}|\frac{\partial^2}{\partial p_k\partial p_k}|{\bf p},\sigma\rangle
+e^{ip\cdot x}\hat{p}_k\frac{\partial}{\partial p_k}\langle 0 |\psi_n(0)|{\bf p},\sigma\rangle\;.
\end{eqnarray}
Using the formula (A.13) for $\phi_{k0}$, we can write 
$$\phi_{k0}x_k e^{ip\cdot x}=\phi_{k0}[x_k+\hat{p}_kx_0] e^{ip\cdot x}=-i\phi_{k0}\frac{\partial}{\partial p_k}e^{ip\cdot x}$$
and put the   fourth term in Eq.~(A.44) in the form
\begin{eqnarray*}
&& 2\sigma\phi_{k0}x_k\langle 0 |\psi_n(x)|{\bf p},\sigma\rangle =-2i\sigma \phi_{k0}\langle 0 |\psi_n(x)\frac{\partial}{\partial p_k}|{\bf p},\sigma\rangle \nonumber \\ &&
+\frac{2\sigma}{|{\bf p}|}\phi_{k0}\sum_m [{\cal J}_{k0}]_{nm}\langle 0 |\psi_m(x)|{\bf p},\sigma\rangle-\frac{2\sigma^2}{|{\bf p}|}\phi_{k0}\phi_{k0}\langle 0 |\psi_n(x)|{\bf p},\sigma\rangle\;.
\end{eqnarray*}
The next-to-last term here cancels the second term in Eq.~(A.44).  Using Eq.~(A.40) again, Eq.~(A.44) now takes almost the desired form:
\begin{eqnarray}
&& \langle 0 |\psi_n(x)K_0|{\bf p},\sigma\rangle=\Bigg\langle 0 \Bigg|\psi_n(x)\Big[|{\bf p}|\frac{\partial^2}{\partial p_k\partial p_k}-2i\sigma\phi_{k0}\frac{\partial}{\partial p_k}\nonumber\\&&
~~~-\frac{\sigma^2}{|{\bf p}|}\phi_{k0}\phi_{k0}+\frac{A+B}{|{\bf p}|}\Big]\Bigg|{\bf p},\sigma\Bigg\rangle\nonumber\\&&~~~
+\frac{1}{|{\bf p}|}\sum_m[{\cal J}_{k0}{\cal J}_{k0}]_{nm}\langle 0 |\psi_m(x)|{\bf p},\sigma\rangle\;.
\end{eqnarray}
Comparing this with Eq.~(A.24), we see that for a field $\psi$ to describe a massless particle of helicity $\sigma$, it is necessary that
\begin{equation}
[{\cal J}_{k0}{\cal J}_{k0}]_{nm}=-\delta_{nm}\Big[\sigma^2+A+B\Big]=-\delta_{nm}\Big[(B-A)^2+A+B\Big]
\end{equation}
But by using Eqs.~(A.36) and (A.38), we see that
\begin{equation}
[{\cal J}_{k0}{\cal J}_{k0}]_{nm}=[{\cal J}_i{\cal J}_i]_{nm}-2\delta_{nm}[A(A+1)+B(B+1)]
\end{equation}
so the requirement (A.46) is that
\begin{equation}
[\BM{\cal J}^2]_{nm}=\delta_{nm}\Big[-(B-A)^2-A-B+2A(
A+1)+2B(B+1)\Big]\;.
\end{equation}
This rules out most irreducible representations of the Lorentz group, for which 
$\BM{\cal J}^2$ takes different values for different components.  For instance, in the $(1/2,1/2)$ four-vector representations, $\BM{\cal J}^2$ takes the value $0$ for the time-component and the value $1(1+1)$ for the space components.  The only irreducible representations for which 
$\BM{\cal J}^2$ takes the same value for all components are the $2j+1$-dimensional representations 
$(j,0)$ and $(0,j)$, with $j$ a positive integer or half-integer.  For all these representations Eq.~(A.48) {\em is} satisfied, since  with either $A=j$ and $B=0$ or $A=0$ and $B=j$, we have $\BM{\cal J}^2=j(j+1)$  and $-(B-A)^2-A-B+2A(
A+1)+2B(B+1)=j(j+1)$.

We conclude then that the only primary field that in a conformally invariant theory can describe a massless particle of helicity $\sigma$  is the $(j,0)$ representation if $\sigma=-j$ or the $(0,j)$ representation if $\sigma=j$.  The conformal dimensionalities of these fields are simply $d=j+1$.
Other fields of type $(A,B)$ with $B-A=\sigma$ can describe a massless particle of helicity $ \sigma=\pm j$, but these are spacetime derivatives of fields of type $(j,0)$ or $(0,j)$, and cannot have the conformal transformation properties (A.30)--(A.33) of a primary field.

\begin{center}
{\bf REFERENCES}
\end{center}

\begin{enumerate}

\item The earliest reference seems to be D. Buchholz and K. Fredenhagen, J. Math. Phys. {\bf 18}, 1107 (1977).  A much simpler and more transparent proof for the case $j=0$ has been given by E. Witten, private communication.  Witten's proof is based on the known explict form of the two-point function for scalar fields in scale-invariant theories.  The proof presented here is a generalization of Witten's proof to $j\neq 0$, but avoids having to work out the explicit form of the two-point function.

\item  G. Mack, Commun. Math. Phys. {\bf 55}, 1 (1977).  Mack's construction of these representation is based on states with simple transformation properties under the compact subgroup $SO(4)\times SO(2)$ of $SO(4,2)$, rather than the states of definite momentum considered here in the appendix, though Mack does identify the mass and helicity of states belonging to various representations of $SO(4)\times SO(2)$.  The massless particle states considered in this paper correspond to item (5) in the table of representations given in Section 1 of Mack's paper.  This representation of the conformal group is sometimes called the ``doubleton'' representation; see, e.g., E. Sezgin and P. Sundell, JHEP {\bf 0109}, 036 (2001) [hep-th/0105001], Section 2.

\item This argument is spelled out  by  P. C. Argyres, M. R. Plesser, N. Seiberg, and E. Witten, Nucl. Phys. B {\bf 461}, 71 (1996).

\item F. Chan and H. F. Jones, Phys. Rev. D {\bf 19}, 1321 (1974).

\item E. P. Wigner, in {\em Theoretical Physics} (International Atomic Energy Agency, Vienna, 1963).  For a textbook treatment, see S. Weinberg, {\em The Quantum Theory of Fields, Vol. I} (Cambridge University Press, Cambridge, UK, 1995), Sec. 2.5.
 
\item S. Weinberg, Phys. Rev. {\bf 134}, B882 (1964).  For a textbook treatment, see S. Weinberg, {\em The Quantum Theory of Fields, Vol. I} (Cambridge University Press, Cambridge, UK, 1995), Sec. 5.9. 
 
\end{enumerate}
\end{document}